\documentstyle[preprint,aps,epsfig]{revtex}
\tightenlines

\newcommand{\lsp}	{{\tilde{\chi}^0_1}}
\newcommand{\es}	{\varepsilon}
\newcommand{\rd}	{\partial}

\begin{document}
\draft
\preprint{
\vbox{\hbox{KIAS-P99005}
      \hbox{SNUTP\hspace*{.2em}98-145}}
}

\title{ Single--photon events in $e^+e^-$ collisions}

\author{
 S.Y.~Choi$^1$, J.S.~Shim$^2$, H.S.~Song$^3$, J.~Song$^3$ and
C.~Yu$^3$ 
}

\address{
$^1$ Korea Institute for Advanced Study, 207--43, Cheongryangri--dong\\
     Dongdaemun--gu, Seoul 130--012, Korea \\
$^2$ Department of Physics, Myongji University, Yongin 449--728,
     Korea \\
$^3$ Center for Theoretical Physics and Department of Physics\\
     Seoul National University, Seoul 151-742, Korea
}

\maketitle
\vspace{2cm}

\begin{abstract}
We provide a detailed investigation of single--photon production processes
in $e^+e^-$ collisions with missing momenta carried 
by neutrinos or neutralinos.
The transition amplitudes for both processes can 
be organized into a generic simplified, factorized form; each neutral 
V$\pm$A vector current of missing energy carriers is factorized out
and all the characteristics of the reaction is solely included in the 
electron vector current. Firstly, we apply the generic 
form to give a unified description of a single--photon production with 
a Dirac--type or Majorana--type neutrino--pair and to confirm their 
identical characteristics as suggested by the so-called Practical 
Dirac--Majorana Confusion Theorem. Secondly, we show that the generic
amplitude form is maintained with the anomalous P-- and C--invariant
WW$\gamma$ couplings in the neutrino--associated process and 
it enables us to easily understand large contributions of the anomalous  
WW$\gamma$ couplings at higher energies and, in particular, at the points 
away from the Z--resonance peak. Finally, the neutralino--associated 
process, which receives modifications in both the left--handed and
right-handed electron currents due to the exchanges of the left--handed 
and right-handed selectrons, can be differentiated from the 
neutrino--associated ones through the left--right asymmetries and/or the 
circular polarization of the outgoing photon.
\end{abstract}
\pacs{\rm PACS number(s):12.15.-y,~12.60.Jv,~14.70.Fm}
%

\section{Introduction}

All the large luminosity and high energy experiments up to now have 
confirmed the validity of the Standard Model (SM) to an unexpectedly high 
level \cite{Altarelli}. In spite of its extraordinary success, the SM has 
a lot of conceptual problems such as the gauge hierarchy problem so that 
it is believed to be valid only at the electroweak scale and to be extended
at higher energies. The first would-be evidence beyond the SM, although 
it has to be independently confirmed by other experiments, has
come from the neutrino sector as the zenith-angle-dependent neutrino 
flux has been observed in the Super--Kamiokande experiment \cite{SuperK}. 
On the other hand,  high energy collider experiments such as LEP2, LHC 
and a high energy $e^+e^-$ linear collider (NLC) should accelerate a broad 
investigation of new physics beyond the SM in the near future.

The process $e^+e^-\rightarrow \gamma+X$ with a distinctive 
``photon--plus--missing--energy'' signal can serve as one of the most 
efficient processes for the exploration of new physics. 
In the process the missing energy can be carried by the SM neutrinos
or weakly interacting or invisible new (s)particles.
In the framework of the SM, the single--photon process with the missing
energy carried by neutrinos has been exploited to count the number 
of light neutrino species at PETRA, SLAC and LEP1 \cite{PETRA,L3}
since, at low energies, the contribution from the $t$--channel W--exchange
diagrams becomes negligible. However, the W--exchange 
contributions become important at high energies
so that the neutrino--associated single--photon process  allows for 
measuring the WW$\gamma$ coupling independently of the WWZ coupling 
unlike the most discussed $e^+e^-\rightarrow W^+W^-$. 

The events with a photon plus missing energy in $e^+e^-$ collisions
might originate from other mechanisms\footnote{Recent developments \cite{RD} 
in superstring theory have led to a radical rethinking of the possibilities
for new particles and dynamics arising from extra compactified 
spatial dimensions. Among the new particle states, the so--called Kaluza--Klein
massive gravitons \cite{KK} can be the invisible particles carrying
the missing energy in the single--photon events.},
signaling new physics beyond the SM. 
For example, such final states can be 
produced in the Minimal Supersymmetric SM (MSSM), one of the most promising 
frameworks for the new theory. The missing energy in these events is caused by 
the weakly interacting or invisible particles such as lightest neutralinos, 
gravitinos and/or sneutrinos. 
In all such cases the SM neutrino-associated single--photon
events are irreducible background. Therefore, in order to reach a 
definite conclusion of new physics, comprehensive calculations
and reliable estimations of all possible single--photon processes are
requisite.

In the present work, we provide a unified description of
the following three cases for single--photon events: 
(i) $e^+ e^-\rightarrow\gamma\nu\nu$ in the SM
including the case when the neutrinos are of Majorana type, (ii) 
$e^+ e^-\rightarrow\gamma\nu\nu$ with the P-- and C--preserving
general WW$\gamma$ coupling, and (iii) $e^+ e^-\rightarrow\gamma\lsp\lsp$
in the MSSM assuming that the
lightest neutralino is the lightest supersymmetric particle (LSP).
Several diagrams are involved in all the processes under consideration
so that the complete calculations look quite demanding. 
However, as will be shown in the following, the transition amplitude of 
every single--photon process is organized into a generic 
simple, unified form; each neutral vector current of missing energy 
carriers is factorized out and all the dynamical 
characteristics for the process are solely included in the electron 
vector current. 

The rest of the present work is organized as follows. 
In Section~II, we exemplify the amplitude reduction procedure for the 
neutrino--associated single--photon process in the SM and apply 
it to give a unified description of a single--photon 
production with a Dirac--type or Majorana--type neutrino--pair, which
facilitate confirming the indistinguishability between the observations
of both processes 
as suggested by the so-called Practical Dirac--Majorana Confusion 
Theorem \cite{Kayser}. Then, we show that the generic amplitude form is 
maintained even after including  P-- and C--preserving anomalous 
WW$\gamma$ couplings in the neutrino--associated process 
and the simplified form clearly exhibits large contributions of the 
anomalous WW$\gamma$ couplings at higher energies and, in particular,
at the points away from the Z--resonance peak. In Section~III we consider
the neutralino--associated single--photon process in the MSSM. 
This process involves modifications in both the left--handed and 
right-handed electron currents due to the left--handed and right-handed 
selectron exchanges. Nevertheless, the simple unified form of the amplitude, 
which appears in the neutrino--associated single--photon process, can 
be also applied to the process with two identical neutralinos of Majorana type 
as final missing--energy states.
Section~IV is devoted to assessing the usefulness of the left--right asymmetry 
and the circular polarization of the outgoing photon in distinguishing the 
neutralino--associated process from the neutrino--associated one.
Finally, we reserve Section~V for the summary and conclusions.

\section{Neutrino--associated Single--photon Processes}
\label{sec:nunu}

\subsection{Amplitude reduction}

In this subsection, we describe how to obtain a simple unified amplitude
for the processes with a distinctive photon--plus--missing-energy through 
the following specific example \cite{Shim}: 
\begin{eqnarray}
\label{eernn}
 e^-(p_1)+ e^+(p_2)\rightarrow\gamma(k)+\nu(k_1)+\bar{\nu}(k_2)
\end{eqnarray}

The neutrino--associated single--photon process (\ref{eernn})
involves five Feynman diagrams in the SM; three W-mediated and 
two Z-mediated ones as shown in Fig.~1.
The application of the Fierz rearrangement formulas
\begin{eqnarray}
\label{Fierz}
\left[\bar{\psi}_1\gamma_\mu P_L\psi_2\right]\,
\left[\bar{\psi}_3\gamma^\mu P_L\psi_4\right]
     &=& 
-\left[\bar{\psi}_1\gamma_\mu P_L\psi_4\right]\,
 \left[\bar{\psi}_3\gamma^\mu P_L\psi_2\right]\,, \nonumber\\
\left[\bar{\psi}_1 P_R\psi_2\right]\,
\left[\bar{\psi}_3 P_L\psi_4\right] 
     &=& \frac{1}{2}
 \left[\bar{\psi}_1\gamma_\mu P_L\psi_4\right]\,
 \left[\bar{\psi}_3\gamma^\mu P_R\psi_2\right]\,,
\end{eqnarray}
to the three W--mediated diagrams reduces the production amplitude 
to a general form
\begin{eqnarray}
\label{amp}
{\mathcal M}&=&\frac{e g^2}{2\cos^2\theta_W}\, 
               \frac{g_{_{ZXX}}}{(k_1 + k_2)^2-m_Z^2}
               \left[\bar{u}(k_1)\gamma_\mu P_L v(k_2)\right]\nonumber\\ 
            && \times \bar{v}_e (p_2)\bigg[
                 \frac{\gamma^\mu(\rlap/\es^* \rlap/k
                      +2p_1\cdot\es^*)}{2p_1\cdot k}
                      \{L_1 P_L+R_1 P_R\} \nonumber\\
            &&\qquad\quad  
              +\frac{(\rlap/\es^*\rlap/k-2p_2\cdot\es^*)\gamma^\mu}{
                       2p_2\cdot k}\{L_2 P_L + R_2 P_R\}\nonumber\\
            &&\qquad\quad 
                +A^{\mu\nu}_L\gamma_\nu P_L
                +A^{\mu\nu}_R\gamma_\nu P_R \,\bigg] u_e(p_1)\,.
\end{eqnarray}
Here, $P_{L,R}=(1\mp\gamma_5)/2$, the parameter $g_{_{ZXX}}$ denotes
the normalized coupling strength of the $ZXX$ vertex 
(e.g. $g_{_{Z\nu\bar{\nu}}}=1$) and 
$\varepsilon^*$ the polarization vector of the outgoing photon.
Since the $We\nu$ vertex is of the left--handed type in the SM,
only the left--handed form factors  are affected by the W--exchange 
diagrams but the right--handed ones are exclusively determined by 
the $Zee$ vertex:
\begin{eqnarray}
\label{SM}
L_i  &=& \epsilon_L + [ 2 p_i \cdot (k+k_i) + m_W^2] f_W\,,
            \quad R_i = \epsilon_R\ \ [i=1,2]\,, \nonumber\\
A^{\mu\nu}_L&=&2 g^{\mu\nu}(k_2-p_1)\cdot\es^* f_W\,, \hskip 1.5cm\quad
            A^{\mu\nu}_R =0\,,
\end{eqnarray}
where $\epsilon_{_L}$ and $\epsilon_{_R}$ are the SM left- and right-handed
couplings for the $Zee$ vertex and $f_W$ is the momentum-dependent 
form factor:
\begin{eqnarray}
\epsilon_{_L}&=& -\frac{1}{2} + \sin^2\theta_W\,, \quad
\epsilon_{_R} =   \sin^2\theta_W \nonumber\\
f_W          &=& -\cos^2\theta_W \frac{2k_1\cdot k_2-m^2_Z}{
                  (2p_1\cdot k_1+m^2_W)(2p_2\cdot k_2+m^2_W)}\,,
\end{eqnarray}
with the electroweak mixing angle $\theta_W$.
Note that the neutrino vector current of the V$-$A form is factored out
and the whole dynamical information of the process is included only 
in the electron vector current. The contributions from the W--mediated 
processes to the form factors vanish at the Z--resonance pole.
The last two terms, $A^{\mu\nu}_L$ and $A^{\mu\nu}_R$, in (\ref{amp}) 
play a role in conserving U(1)$_{\rm EM}$ gauge invariance and they are
proportional to the factor $f_W$. 

The expression in eq.~(\ref{amp}) is of a very generic form so that 
it can be applied to the amplitude for any process 
producing a single photon 
and a fermion--pair in $e^+e^-$ collisions. This property will be
explicitly demonstrated with three examples; (i) the production of
a photon and a Majorana neutrino pair, (ii) the case with the anomalous
WW$\gamma$ couplings and (iii) the production of a photon and a lightest
neutralino pair. 

In order to check the validity of the simplified form for
the process $e^+e^-\rightarrow\gamma\nu\bar{\nu}$, we 
perform a Monte--Carlo phase--space integration by BASES \cite{BASES} 
with the expression in eq.~(\ref{amp}) and illustrate in Fig.~2 the 
dependence of the differential cross section on the photon energy 
fraction $x_\gamma$ with respect to the electron beam energy $E_b$ 
[$=\sqrt{s}/2$].
Numerically, we find that the differential cross section is completely
consistent with that in Ref.~\cite{Bento}. As can be easily checked from
the simplified form of the amplitude, the peaks in the differential 
cross section ${\rm d}\sigma/{\rm d}x_\gamma$ are attributed to the 
$s$-channel Z-mediated diagrams near the photon energy fraction 
$x_\gamma=1-m_Z^2/s$.

\subsection{Dirac versus Majorana}

In the SM, only the neutrinos among fundamental fermions may 
possess no global discrete quantum numbers such as the lepton numbers, 
opening the possibility that neutrinos are 
their own anti-particles, that is to say, Majorana particles. 
In the light of this aspect, whether light neutrinos are Dirac or Majorana 
particles has been one of the main issues in neutrino physics.
The answer is truly meaningful only when any difference is experimentally 
observed. In the wide range of neutrino experiments at the colliders,
the so--called ``Practical Dirac-Majorana Confusion (PDMC) Theorem"  
in Ref.~\cite{Kayser} holds true \cite{Zralek}. Related with the recent 
evidence of neutrino oscillation, it will be of particular interest to 
check the possibility of determining in the neutrino--associated 
single--photon process whether the produced neutrinos are of Dirac or 
Majorana type or not.

In principle, there exist some differences at the amplitude level
due to different Feynman rules for both types of neutrinos \cite{Denner}.
Compared to Dirac particles, Majorana particles can exhibit two important 
characteristic features: lepton-number violation and different Feynman 
rules for interaction vertices involving the Majorana
particles. In the reaction $e^+e^-\rightarrow\gamma\nu\nu$ 
for a Majorana neutrino pair, there exists a $u$-channel lepton--number 
violating diagram corresponding to each $t$-channel lepton--number preserving 
diagram. 
Due to the fact that there is no vector current for Majorana fermions, the
neutral vector current must be of the type $(\gamma^\mu P_L-\gamma^\mu P_R)$ 
while the charged vector current remains intact. Nevertheless, we will show 
that, if the neutrinos are not detected and (almost) massless,  
the experimental signatures at high--energy colliders are identical 
for both Dirac and Majorana neutrinos. 
This is an additional demonstration of the PDMC theorem. 

For Majorana neutrinos, the amplitude of each $u$-channel diagram is related 
to that of corresponding $t$-channel one by
\begin{eqnarray}
\label{ut}
{\mathcal M}_u(k_1,k_2) = -{\mathcal M}_t (k_2,k_1)\,,
\end{eqnarray}
where the minus sign stems from the interchange of two identical fermions.
On the other hand, the neutral vector current of Majorana neutrinos in
the Z--mediated diagrams can be expressed in terms of two Dirac--type 
amplitudes by
\begin{eqnarray}
\label{Z-Maj-amp}
\bar{u}_M(k_1)\left(\gamma^\mu P_L-\gamma^\mu P_R\right)v_M(k_2)
    = \bar{u}_M(k_1)\gamma^\mu P_Lv_M(k_2) 
     -\bar{u}_M(k_2)\gamma^\mu P_L v_M(k_1) \,,
\end{eqnarray}
where we have used Majorana conditions
\begin{eqnarray}
\label{M-condition}
\bar{u}_M(k_1)\gamma^\mu(1\pm\gamma_5 ) v_M(k_2)
    =
\bar{u}_M(k_2)\gamma^\mu(1\mp\gamma_5 ) v_M(k_1)\,.
\end{eqnarray}
As a result, the production amplitude for Majorana neutrinos is
expressed by 
\begin{eqnarray}
\label{Majorana-Dirac}
{\mathcal M}_M ={\mathcal M}_D(k_1,k_2)-{\mathcal M}_D(k_2,k_1)\,.
\end{eqnarray}
Note that the second term in the right hand side of (\ref{Majorana-Dirac})
is the negative of the first term with $k_1^\mu$ and $k_2^\mu$
exchanged. Therefore, the transition amplitude is expressed in terms
of two amplitudes which are of the generic form in eq.~(\ref{amp}).

We first note that the interference term 
${\mathcal M}_D(k_1,k_2)^* {\mathcal M}_D(k_2,k_1)$ in the evaluation of 
$|{\mathcal M}_M|^2$ becomes, with the help of the expression (\ref{amp}) 
and the Majorana condition (\ref{M-condition}),
\begin{eqnarray}
\label{cross}
{\mathcal M}_D(k_1,k_2)^* {\mathcal M}_D(k_2,k_1)
    = {\mathcal E}_{\mu\nu} \sum_{\rm spin} 
\bar{v} (k_1) \gamma^\mu P_L u(k_2) \,
\bar{u}(k_2) \gamma^\nu P_R v(k_1)
    =
2 m_\nu^2 \, g^{\mu\nu} \, {\mathcal E}_{\mu\nu}\,.
\end{eqnarray}
where ${\mathcal E}_{\mu\nu}$ is a covariant tensor composed of the
absolute square of the electron vector current. The final term in
eq.~(\ref{cross}) is obtained by assuming a finite neutrino mass $m_\nu$
and taking the polarization sum.
Clearly, when the neutrino mass is negligible compared to the beam
energy, the contribution from the interference term vanishes.
The practical incapability of explicitly identifying neutrinos
at high--energy collider experiments forces us to integrate the 
differential cross section over the final phase space of neutrinos.  
As a result, the complete identity of the integrals of two 
squared amplitudes over the symmetric phase space of the two neutrino momenta
\begin{eqnarray}
\int{\rm d}\Phi_2\,\delta^4(k_1 + k_2 -q)
    \left|{\mathcal M}_D(k_1,k_2)\right|^2
  = \int{\rm d}\Phi_2\,\delta^4(k_1+k_2 -q) 
    \left| {\mathcal M}_D(k_2,k_1)\right|^2 \,,
\end{eqnarray}
does not leave any difference in the observation of Dirac and Majorana 
neutrinos.

In summary, any practically observable difference between Dirac and 
Majorana neutrinos can appear only when neutrinos have a 
non--negligible mass.

\subsection{Anomalous WW$\gamma$ coupling}

Under the assumption that the discrete symmetries P, C, and T 
are preserved separately, the general coupling of two charged vector bosons
W$^\pm$ with a photon $\gamma$ is derived from the most general and 
U(1)$_{\rm EM}$ gauge-invariant Lagrangian \cite{Hagiwara}
\begin{eqnarray}
\label{L-WWgamma}
\frac{{\mathcal L}_{WW\gamma}}{g}
   =
  i\,( W^\dagger_{\mu\nu} W^\mu A^\nu -W^\dagger_\mu A_\nu W^{\mu\nu})
+i\,\kappa_\gamma\,W^\dagger_\mu W_\nu F^{\mu\nu} 
+\frac{i\lambda_\gamma}{m_W^2} W^\dagger_{\lambda\nu}
  W^\mu_\nu F^{\mu\lambda} \,,
\end{eqnarray}
where $W^{\mu\nu}={\rd}^\mu W^\nu-{\rd}^\nu W^\mu$ and
$F^{\mu\nu}=\partial^\mu A^\nu-\partial^\nu A^\mu$.
The parameters $\kappa_\gamma$ and $\lambda_\gamma$, [which are 1 and 0
in the SM], are related to the anomalous magnetic dipole moment $\mu_W$
and the electric quadrupole moment $Q_W$ of the W boson by
\begin{eqnarray}
\mu_W &=& \frac{e (1+\kappa_\gamma +\lambda_\gamma)}{2 m_W}\,, \nonumber\\ 
Q_W   &=&-\frac{e (\kappa_\gamma-\lambda_\gamma)}{m_W^2}\,.
\end{eqnarray}

These self-interactions of gauge bosons have been extensively investigated 
through various processes at $e^+ e^-$ and hadron colliders \cite{W-previous}.
Among them the hadron-free reaction $e^+e^-\rightarrow\gamma\nu\bar{\nu}$ is 
favorable in the investigation of the WW$\gamma$ vertex since it does not 
include the other self interactions of gauge bosons \cite{JKW}.
Even though the gauge group structure of the SM specifies
the self interactions of the W, Z and $\gamma$ when regarded as fundamental 
gauge bosons, their precise confirmation is to be experimentally  
established \cite{PDG}. The ALEPH collaboration has reported preliminary 
results for the coupling $\kappa_\gamma -1=0.05^{+1.2}_{-1.1}$ (stat.) and 
$\lambda_\gamma=-0.05^{+1.6}_{-1.5}$ (stat.) from the data of the process 
$e^+e^-\rightarrow\gamma\nu\bar{\nu}$ at $\sqrt{s}=161$, 172, and 
183 GeV \cite{ALEPH}. Any deviation from the SM prediction will lead to 
the hint for the theory beyond the SM. 

After a little lengthy calculation, we find that even in the existence of
the anomalous WW$\gamma$ couplings the transition amplitude  
for this reaction still keeps the unified form (\ref{amp}) with the 
following modifications in the form factors:
\begin{eqnarray}
\label{general-coupling}
L_i &=& \epsilon_L 
      + \left[ \left(1+\kappa_\gamma
      -2\frac{\lambda_\gamma}{m^2_W}(p_i\cdot k_i)\right)(p_i\cdot k)
           +2p_i\cdot k_i + m_W^2 \right] f_W\,, \nonumber\\
R_i &=& \epsilon_R\,, \nonumber\\
A^{\mu\nu}_L &=& \left[ \,g^{\mu\nu} \es^* \cdot (K_2-K_1) 
          + 2\frac{\lambda_\gamma}{m^2_W}(p_1-k_1)\cdot k {\es^*}^\mu k^\nu 
          - 2\frac{\lambda_\gamma}{m^2_W}(p_1-k_1)\cdot\es^* 
          k^\mu k^\nu\,\right]f_W\,,
          \nonumber\\ 
A^{\mu\nu}_R &=& 0 \,, \nonumber\\  
K_i    &=& k_i+\kappa_\gamma\, p_i -2\frac{\lambda_\gamma}{m^2_W}
           (p_i\cdot k_i) \, p_i\,.
\end{eqnarray}
Compared to the transition amplitude in the SM, only the V$-$A part of the
electron vector current is modified. This is understandable because 
the V$-$A vertex remains the same for the charged electron current with 
the W boson which is to be coupled with the photon.
In Fig.~3, we show the differential cross section with respect to 
the photon energy fraction $x_\gamma$ at $\sqrt{s}=200$ GeV and 
$\sqrt{s}=500$ GeV for two cases\footnote{The conservative 
ranges of the parameters $\kappa_\gamma$ 
and $\lambda_\gamma$ quoted in Ref.~\cite{PDG} are considered.}: 
three values of $\kappa_\gamma$ with 
$\lambda_\gamma=0$ [$\kappa_\gamma=1,-1.3$ and 3.2] and three values of 
$\lambda_\gamma$ with $\kappa_\gamma=1$ [$\lambda_\gamma=0,-1$ and 1].
Note that the effects of non-standard couplings increase 
at higher energies \cite{Bilchak}, reflecting the fact that the anomalous 
terms are higher-dimensional and non--renormalizable.
The figure clearly shows that it is very difficult to observe the deviations
due to the anomalous parameters $\kappa_\gamma$ and $\lambda_\gamma$ near 
$x_\gamma=1-m^2_Z/s$ where the Z--exchange contributions dominate over the 
W--exchange ones. Therefore, it is crucial to apply appropriate
photon energy cuts to enhance the possibility to see the anomalous effects.

\section{Neutralino--associated Single--photon Process} 

Supersymmetry is a new symmetry which provides a well--motivated 
extension of the SM with an elegant solution to the gauge hierarchy
problem.  Most supersymmetry theories assume the so--called R-parity 
under which the SM particles are even and the supersymmetric particles 
are odd. The conservation of R--parity ensures
the stability of the LSP so that it escapes from the detection. 
In most supersymmetric models, the lightest neutralino $\tilde{\chi}^0_1$ 
is the LSP in a wide range of parameter space.  
Because of the elusive property, the existence of the lightest 
neutralino can not be checked through the simplest process 
$e^+e^-\rightarrow \tilde{\chi}^0_1\tilde{\chi}^0_1$ leaving
no signals in a detector. However, the production of a lightest 
neutralino pair accompanied by a single photon in $e^+e^-$ collisions
can give useful information on the existence of the LSP
through the photon energy and angular distributions along with
tuning the electron beam polarization and/or measuring the outgoing  
photon polarization. 

In this section, we concentrate on the single--photon process
$e^+e^-\rightarrow\gamma\tilde{\chi}^0_1\tilde{\chi}^0_1$ in the MSSM.
Because of the electroweak gauge symmetry breaking, the
gauginos, the superpartners of gauge bosons, and the higgsinos, the
superpartners of the Higgs bosons, can mix to give physical mass 
eigenstates in the MSSM. 
In particular, the photino $\tilde{\gamma}$ and the Zino
$\tilde{Z}$ mix with two neutral higgsinos $\tilde{H}^0_1$ and
$\tilde{H}^0_2$ to form four neutralino mass eigenstates 
$\tilde{\chi}^0_i$ [$i=1$ to 4]. The neutralino masses and 
the mixing angles are determined by $m_Z$, $\tan\beta$, two
soft SUSY--breaking gaugino mass parameters $M_1$ and $M_2$ and
the SUSY--preserving higgsino mass parameter $\mu$. The symmetric
$4\times 4$ neutralino mass matrix can be diagonalized by a $4\times 4$
unitary matrix $N$ \cite{Haber}.  Despite the involved neutralino mixing
as well as the large number of Feynman diagrams,
we will show that the production amplitude for the process
$e^+e^-\rightarrow\gamma\tilde{\chi}^0_1\tilde{\chi}^0_1$ can be also  
organized into the unified form in eq.~(\ref{amp}),
which enables us to investigate the dependence of the
energy and angular spectrum of the outgoing photon on the relevant SUSY 
parameters.

The reaction $e^+e^-\rightarrow\gamma\tilde{\chi}^0_1\tilde{\chi}^0_1$
in the MSSM involves 14 Feynman diagrams as depicted in Fig.~4.
The selectron-exchange diagrams with the primed indices
[Figs.~(c')-(h')] are allowed due to the Majorana property of neutralinos, 
of which the amplitudes are related to those of the corresponding 
$t$-channel ones by
\begin{eqnarray}
{\mathcal M}_x'(k_1, k_2) = -{\mathcal M}_x(k_2,k_1) \quad 
                        [x =  c,d,e,f,g,h]\,,
\end{eqnarray}
where $k_1$ and $k_2$ are the four-momenta of the two lightest
neutralinos. Due to the Majorana condition in eq.~(\ref{M-condition}) 
the diagrams (A) and (B) can be expressed by 
\begin{eqnarray}
{\mathcal M}_{A,B}={\mathcal M}_{a,b}(k_1, k_2)-{\mathcal M}_{a,b}(k_2,k_1)
                 \equiv {\mathcal M}_{a,b} (k_1, k_2) 
                  + {\mathcal M}_{a,b}' (k_1, k_2) \,.
\end{eqnarray}
Defining the following combination to be ${\mathcal M}_L$:
\begin{eqnarray}
{\mathcal M}_L \equiv {\mathcal M}_a+{\mathcal M}_b
                    + {\mathcal M}_{c}'+{\mathcal M}_{d}'
                    +{\mathcal M}_{e}'+{\mathcal M}_f
                    +{\mathcal M}_g+{\mathcal M}_h\,,
\end{eqnarray}
we can show that the sum of the remaining amplitudes,
denoted by ${\mathcal M}_R$,
satisfies the relation
\begin{eqnarray}
{\mathcal M}_R(k_1, k_2) = - {\mathcal M}_L(k_2,k_1) \,,
\end{eqnarray}
and thus the total production amplitude ${\mathcal M}$
for the reaction $e^+e^-\rightarrow\gamma\tilde{\chi}^0_1\tilde{\chi}^0_1$
is given by
\begin{eqnarray}
{\mathcal M} = {\mathcal M}_L+ {\mathcal M}_R 
             = {\mathcal M}_L(k_1, k_2)- {\mathcal M}_L (k_2,k_1)\,.
\end{eqnarray}
Then, the Fierz rearrangement formulas in eq.~(\ref{Fierz})
cast the production amplitude into the unified form in eq.~(\ref{amp}) with 
the following modifications:
\begin{eqnarray}
g_{_{Z\lsp\lsp}}&=&\frac{1}{2}\left[|N_{13}|^2-|N_{14}|^2\right] \,,
\nonumber\\
L_1 &=& \epsilon_{_L}-\frac{1}{2}\left[(p_1-k_2)^2 
          -m^2_{\tilde{e}_{L}}\right]f_{\tilde{e}_{L}}\,, \quad
L_2  =  \epsilon_{_L}-\frac{1}{2}\left[(p_2-k_1)^2
          -m^2_{\tilde{e}_{L}}\right]f_{\tilde{e}_{L}}\,, \nonumber\\ 
R_1 &=& \epsilon_{_R} + \frac{1}{2}\left[(p_1-k_1)^2 
          -m^2_{\tilde{e}_{R}}\right]f_{\tilde{e}_{R}}\,, \quad
R_2  = \epsilon_{_R} + \frac{1}{2}\left[(p_2-k_2)^2 
          -m^2_{\tilde{e}_{R}}\right]f_{\tilde{e}_{R}}\,, \nonumber\\
A_L^{\mu\nu} &=& g^{\mu\nu}(k_2-p_1)\cdot \es^* f_{\tilde{e}_{L}}\,, \quad
A_R^{\mu\nu}  =  g^{\mu\nu} (k_2-p_2)\cdot \es^* f_{\tilde{e}_{R}}
\end{eqnarray}
where the form factors $f_{\tilde{e}_L}$ and $f_{\tilde{e}_R}$ describing 
the selectron-exchanges are given by
\begin{eqnarray}
f_{\tilde{e}_{L}}&=&\frac{4\cos^2\theta_W | g_{L}|^2}{ g_{Z\lsp\lsp} }
                    \frac{(k_1+k_2)^2 -m_Z^2}{[(p_1-k_2)^2
                    -m^2_{\tilde{e}_{L}}][(p_2-k_1)^2-m^2_{\tilde{e}_{L}}]}\,,
                      \nonumber\\
f_{\tilde{e}_{R}}&=&\frac{4\cos^2\theta_W | g_{R}|^2}{ g_{Z\lsp\lsp} }
                    \frac{ (k_1+k_2)^2-m_Z^2}{[(p_1-k_1)^2
                    -m^2_{\tilde{e}_{R}}][(p_2-k_2)^2-m^2_{\tilde{e}_{R}}]}\,,
\end{eqnarray}
with $g_L=\left(N_{12}+\tan\theta_W N_{11}\right)/2$ and 
$g_R=\tan\theta_W N_{11}$.
The factorization of the neutral vector currents of invisible neutralinos 
occurs again at the amplitude level. Compared to the amplitudes 
of the neutrino-associated processes in (\ref{SM}) and 
(\ref{general-coupling}), we observe that the V$+$A structure of the electron
current undergoes considerable changes due to the existence of the 
right--handed selectron exchanges. 
As a result, the use of the right--handed electron beam may be
very helpful to reduce the SM background effects.
This feature will be quantitatively demonstrated in the next section.

In Fig.~5, we have demonstrated the differential cross section
with respect to the photon energy fraction $x_\gamma$ at $\sqrt{s}=200$ GeV
and 500 GeV for $\tan\beta=2$ and 30, respectively, taking 
$m_{\tilde{e}_{L,R}}=100$ GeV.
The lightest neutralino mass and the elements of the mixing matrix $N$
are computed by using $M_1=100$ GeV, $\mu =100$ GeV, and  the 
assumption of the gaugino mass unification condition 
$M_1= (5/3)\tan^2\theta_W M_2$.
We note that for $\tan\beta=30$ the resonance peak around the Z--resonance
pole is absent which is apparently present for $\tan\beta=2$.
These different behaviors according to $\tan\beta$ can be
explained by comparing the maximally allowed photon 
energy $x^{\rm max}_\gamma$
with the photon energy fraction for the resonance 
peak $x^{\rm Z-peak}_\gamma$. 
The maximum energy fraction of the photon corresponds to the largest momentum
which is obtained when the photon is scattered against the collinear
neutralinos: 
\begin{eqnarray}
\label{max-energy}
x^{\rm max}_\gamma = 1-\frac{4 m_{\lsp}^2}{s}\,.
\end{eqnarray}
Since the resonance peak occurs at $x^{\rm Z-peak}_\gamma =1-m_Z^2/s$,
there is no peak if $m_{\lsp}\geq m_Z/2$. With the above numerical 
values for $M_1$, $M_2$, $\mu$, we have $m_{\lsp} =39$ GeV for $\tan\beta=2$ 
and $m_{\lsp} =61$ GeV for $\tan\beta=30$, 
which correctly explains the different 
behaviors. Therefore, a precise confirmation of the existence of the 
resonance peak after subtracting the SM background effects can provide
valuable information on the lightest neutralino mass $m_{\tilde{\chi}^0_1}$
in the process $e^+e^-\rightarrow\gamma\tilde{\chi}_1^0\tilde{\chi}_1^0$.

\section{Left-right Asymmetries and Photon Polarization}
\label{sec:pol}

One crucial difference of the neutralino--associated  process from
the neutralino--associated one is the existence of the right--handed
selectron--exchanges, so that the ratio of the production rate with
the right--handed electron beam to that with
the left--handed one can be substantially large. 
Since a highly polarized electron beam with its beam polarization more than 
90\% is expected at future $e^+e^-$ linear colliders \cite{FLC}, 
it will be
valuable to study the left--right asymmetries in identifying  the
origin of the single--photon events. Moreover, it is expected that the
circular polarization of the outgoing photon is different.
In this light, we present 
a quantitative analysis for the left--right asymmetries and the
photon circular polarization in the single--photon processes
$e^+e^-\rightarrow\gamma\nu\bar{\nu}$ and 
$e^+e^-\rightarrow\gamma\tilde{\chi}^0_1\tilde{\chi}^0_1$. 

In order to measure a left--right asymmetry $A_{LR}$ defined by
\begin{eqnarray}
A_{LR}=\frac{\sigma_R-\sigma_L}{\sigma_R+\sigma_L}
\,,
\end{eqnarray}
we have only to switch the longitudinal electron polarization, which
should be straightforward in a $e^+e^-$ linear collider. 
In order to measure the circular polarization of the 
final photon beam, we use a general method \cite{Choi} which can be 
applied to any process producing a single photon. 
In the general formalism, the circular polarization is described by 
a Stokes' parameter $\xi_2$ that is nothing but the rate asymmetry:
\begin{eqnarray}
\xi_2=\frac{N_+-N_-}{N_++N_-}
\,,
\end{eqnarray}
where $N_\pm$ is the number of produced photons with positive and
negative helicities.

Figure~6 shows the left--right asymmetries $A_{LR}$ 
as a function of the photon scattering angle $\theta$ 
at $\sqrt{s}=200$ and 500 GeV
with the same SUSY parameters as in Fig.~5. 
The upper frame in the figure is 
for the neutrino--associated process, 
while the middle and lower frames are 
for the neutralino--associated ones for $\tan\beta=2$ [middle] and 
$\tan\beta=30$ [lower]. 
Clearly, the left--right asymmetries are very 
different in two processes; the asymmetries for the neutralino--associated
process are always larger and even positive for $\sqrt{s}=500$ GeV.
Moreover, the dependence of the asymmetry
on $\tan\beta$ becomes significant at $\sqrt{s}=200$ GeV.
As discussed in the previous section, 
the right--handed electron beam is very useful to 
identifying the neutralino--associated process, removing the large 
portion of the SM background.

In Fig.~7 we show the circular polarization degree $\xi_2$ of the outgoing 
photon as a function of the photon scattering angle $\theta$ for the 
neutrino--associated process [(a) and (c)] and the neutralino--associated 
one [(b) and (d)] with the same SUSY parameters as in Figs.~5 and 6.
We set the electron beam to be purely left--handed in (a) and (b) and 
right--handed in (c) and (d). For the left--handed [right--handed] electron 
beam, $\xi_2$ is negative [positive] in the forward direction and 
positive [negative] in the backward direction, respectively.
Note that the circular polarization in the neutralino--associated  process
is more sensitive to the beam energy of the right--handed electron beam
than of the left--handed electron beam. This dependence is, however, opposite
in the neutrino--associated process. 

\section{Conclusions}
\label{sec:con}

We have studied in detail the single--photon events in high--energy
$e^+ e^- $ collisions as attributing the missing energy to neutrinos 
in the SM including the effects of the anomalous WW$\gamma$ couplings, 
or to neutralinos in the MSSM, which are assumed to be the LSP.
We have found that the transition amplitudes for both processes can 
be organized into a generic simplified, factorized form; each neutral 
V$\pm$A vector current of missing energy carriers is factorized out
and all the characteristics for the reaction is solely included in the 
electron vector current. 

The amplitude reduction procedure described in Section~II.A allows us 
to give 
a unified description of a single--photon production with 
a Dirac--type or Majorana--type neutrino--pair and to easily confirm 
their identical characteristics in the observation supported by the 
so-called Practical Dirac--Majorana Confusion Theorem. 
The generic amplitude form is preserved
with the anomalous WW$\gamma$ couplings in the neutrino--associated 
process and it enables us to easily understand large contributions of 
the anomalous P-- and  C--invariant WW$\gamma$ couplings at higher energies 
and, in particular, at the points away from the Z peak. 
The neutralino--associated single--photon process in the MSSM
involves the modification in both the left--handed and right-handed 
electron currents due to the left--handed and right-handed selectron 
exchanges. Nevertheless, the basic simplified amplitude form can 
be applied to the production process 
of two identical neutralinos of Majorana type as well. 
We have found that, due to these distinct properties, utilizing the 
left--right asymmetries for the longitudinal electron polarization and/or 
measuring the circular polarization of the outgoing photon may be very 
useful in disentangling the neutralino--associated processes from 
the neutrino--associated ones.

\acknowledgments

HS and CY would like to acknowledge the financial support of the
Korea Research Foundation through the 97 Sughak Program and
98--015-D00054. The work of JHS was supported by the Korea Science 
and Engineering Foundation (KOSEF) through the Center for Theoretical Physics,
Seoul National University and the work of SYC supported by 
KOSEF through the KOSEF--DFG large collaboration project, 
Project No. 96--0702--01--01-2.

\bigskip
\bigskip

\newpage
\begin{center}
\begin{large}
FIGURE CAPTIONS
\end{large}
\vspace{5mm}\
\end{center}
\begin{description}
\item [Fig.~1]
Feynman Diagrams contributing to the neutrino--associated
single--photon process $e^+e^-\rightarrow\nu\bar{\nu}\gamma$
in the SM.

\item [Fig.~2]
The differential cross section of the neutrino--associated
process $e^+e^-\gamma\rightarrow\nu\bar{\nu}$ with respect
to the energy fraction of the outgoing photon $x_\gamma$.
The solid line is for $\sqrt{s}=200$ GeV and the dotted
line for $\sqrt{s}=500$ GeV.

\item [Fig.~3]
The photon energy distributions ${\rm d}\sigma/{\rm d}x_\gamma$
of the process $e^+e^-\rightarrow\gamma\nu\bar{\nu}$ for the
$e^+e^-$ c.m. energy of 200 GeV and 500 GeV with
the different values for the anomalous parameters,
$\kappa_\gamma$ and $\lambda_\gamma$. In (a) and (c) the
value of $\lambda_\gamma$ is taken to be 0 and in (b) and (d)
the value of $\kappa_\gamma$ taken to be 1.

\item [Fig.~4]
Feynman diagrams contributing to the neutralino--associated
single--photon process $e^+e^-\rightarrow\gamma\tilde{\chi}^0_1
\tilde{\chi}^0_1$ in the MSSM.

\item [Fig.~5]
The different cross section of the process
$e^+e^-\rightarrow\gamma\tilde{\chi}^0_1\tilde{\chi}^0_1$
with respect to the photon energy fraction $x_\gamma$.
The lightest neutralino mass and
its couplings are calculated with $M_1=100$ GeV, $\mu=100$ GeV
and the gaugino unification condition
$M_1=(5/3)\tan^2\theta_W M_2$ for two $\tan\beta$ values;
$\tan\beta=2$ in the left frame and
$\tan\beta=30$ in the right frame.
The masses for the right--handed and left--handed selectrons
are assumed to be 100 GeV.

\item [Fig.~6]
The left--right asymmetries as a function of the photon
scattering angle $\theta$ at $\sqrt{s}=200$ and 500 GeV.
The upper frame is for the neutrino--associated process,
while the middle and lower frames are for the
neutralino--associated processes for $\tan\beta=2$ [middle]
and $\tan\beta=30$ [lower]. The other SUSY parameters are taken
to be the same as in Fig.~5.

\item [Fig.~7]
The degree of photon circular polarization as a function
             of the photon scattering angle $\theta$ for the
             neutrino--associated process [(a) and (c)] and
             the neutralino--associated process [(b) and (d)].
             The electron beam is purely left--handed in (a) and (b),
             and right--handed in (c) and (d). The other SUSY parameters
             have the same values as in Fig.~5.
\end{description}

%
\begin{center}
\begin{figure}[htb]
\vspace{7cm}
\hbox to\textwidth{\hss\epsfig{file=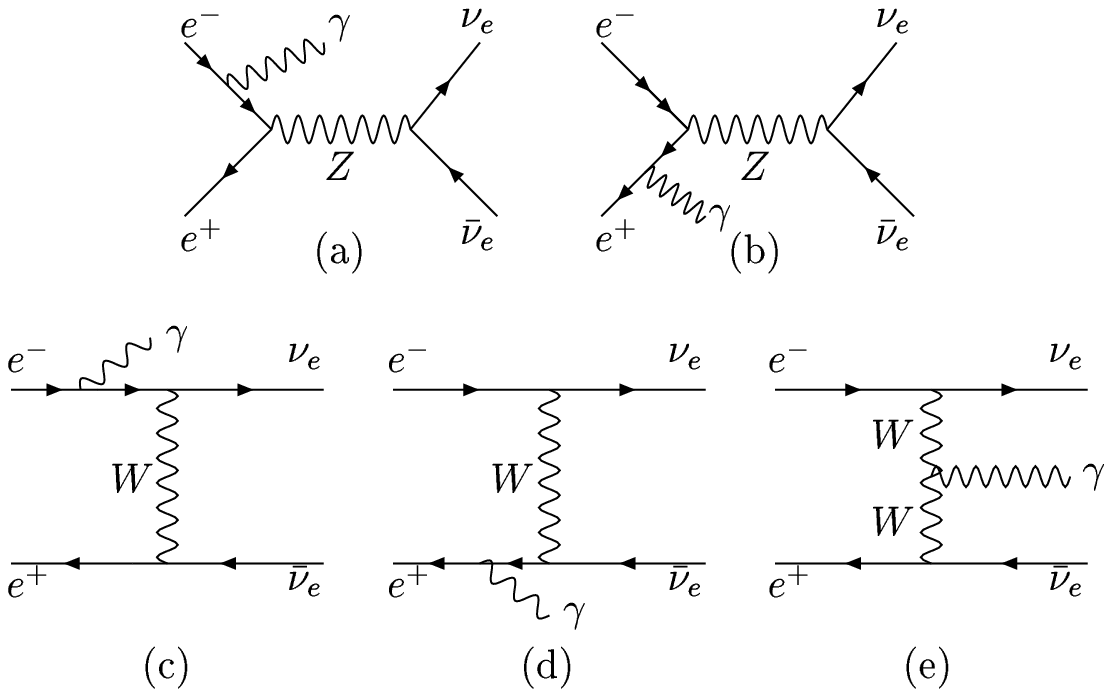,width=11cm,height=7cm}\hss}
\label{fig:fig1}
\end{figure}
\end{center}
\vfill
\begin{center}
{\bf\large Figure 1}
\end{center}

\newpage
\mbox{ }

\begin{center}
\begin{figure}[htb]
\vspace{3cm}
\hbox to\textwidth{\hss\epsfig{file=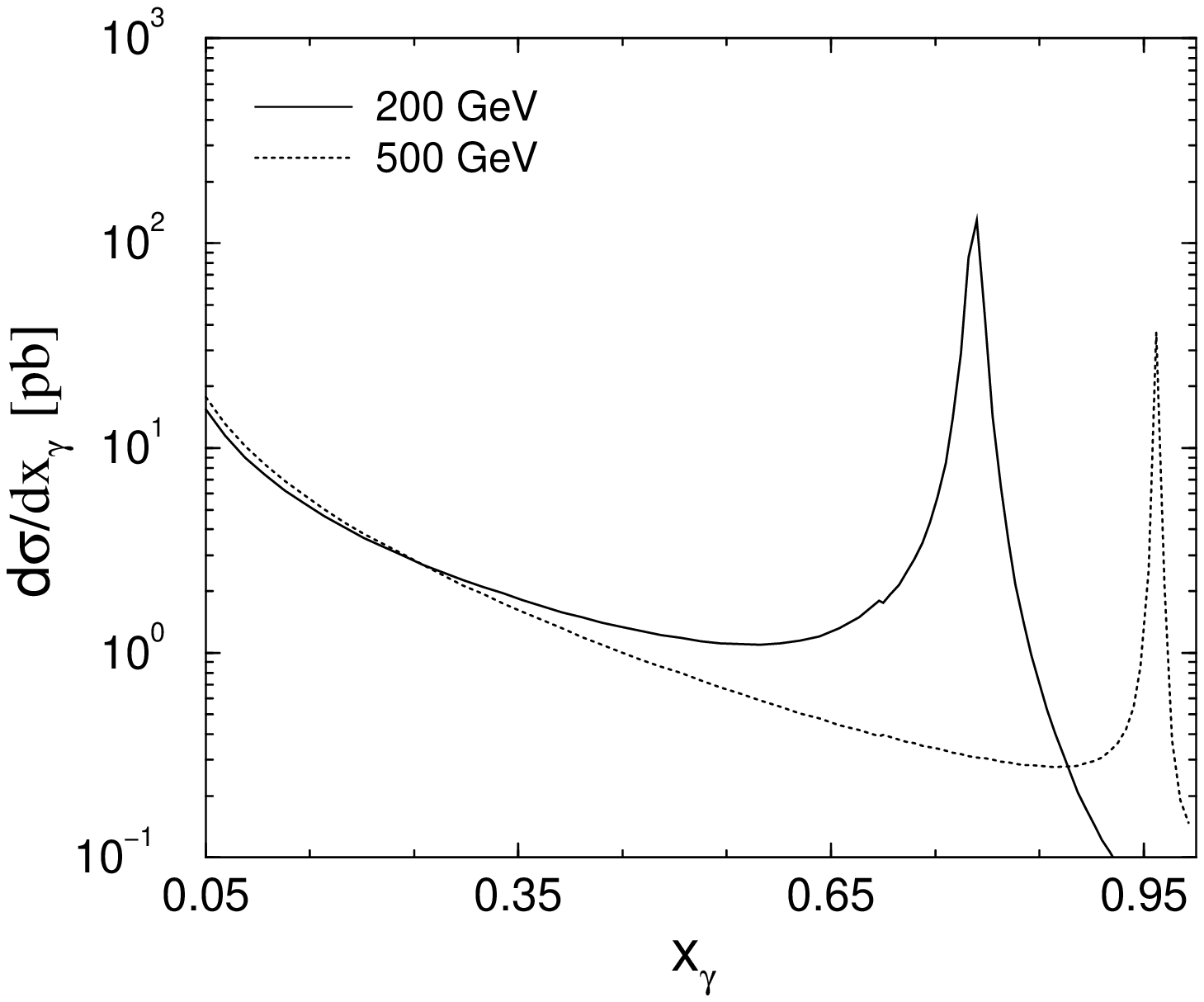,width=13cm,height=13cm}\hss}
\label{fig:fig2}
\end{figure}
\end{center}
\vfill
\begin{center}
{\bf\large Figure 2}
\end{center}

\newpage
\mbox{ }

\begin{center}
\begin{figure}[htb]
\vspace{3cm}
\hbox to\textwidth{\hss\epsfig{file=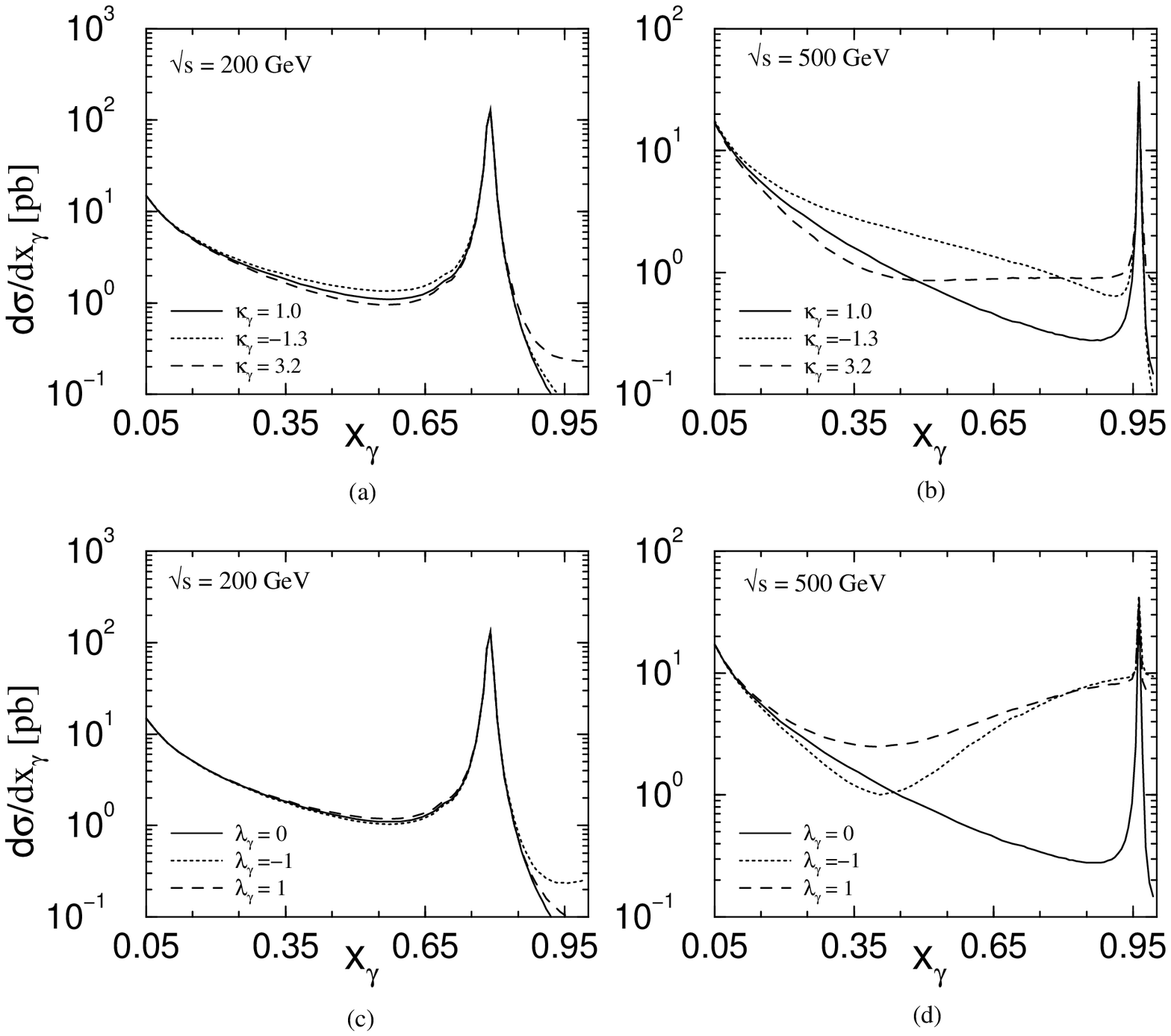,width=12cm,height=13cm}\hss}
\label{fig:fig3}
\end{figure}
\end{center}
\vfill
\begin{center}
{\bf\large Figure 3}
\end{center}

\newpage
\mbox{ }

\begin{center}
\begin{figure}[htb]
\hbox to\textwidth{\hss\epsfig{file=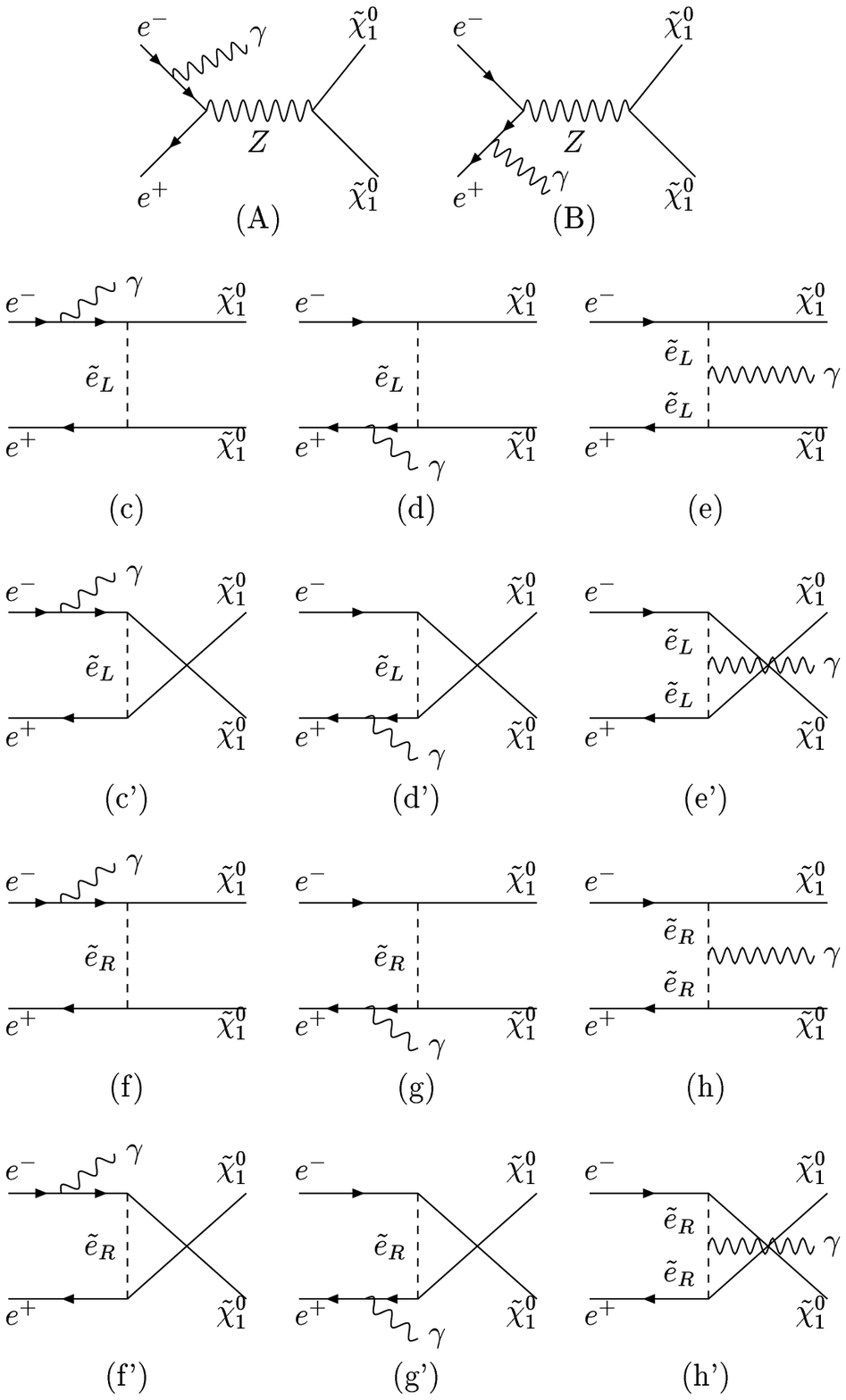,width=11cm,height=19cm}\hss}
\label{fig:fig4}
\end{figure}
\end{center}
\vfill
\begin{center}
{\bf\large Figure 4}
\end{center}

\newpage
\mbox{ }

\begin{center}
\begin{figure}[htb]
\vspace{4cm}
\hbox to\textwidth{\hss\epsfig{file=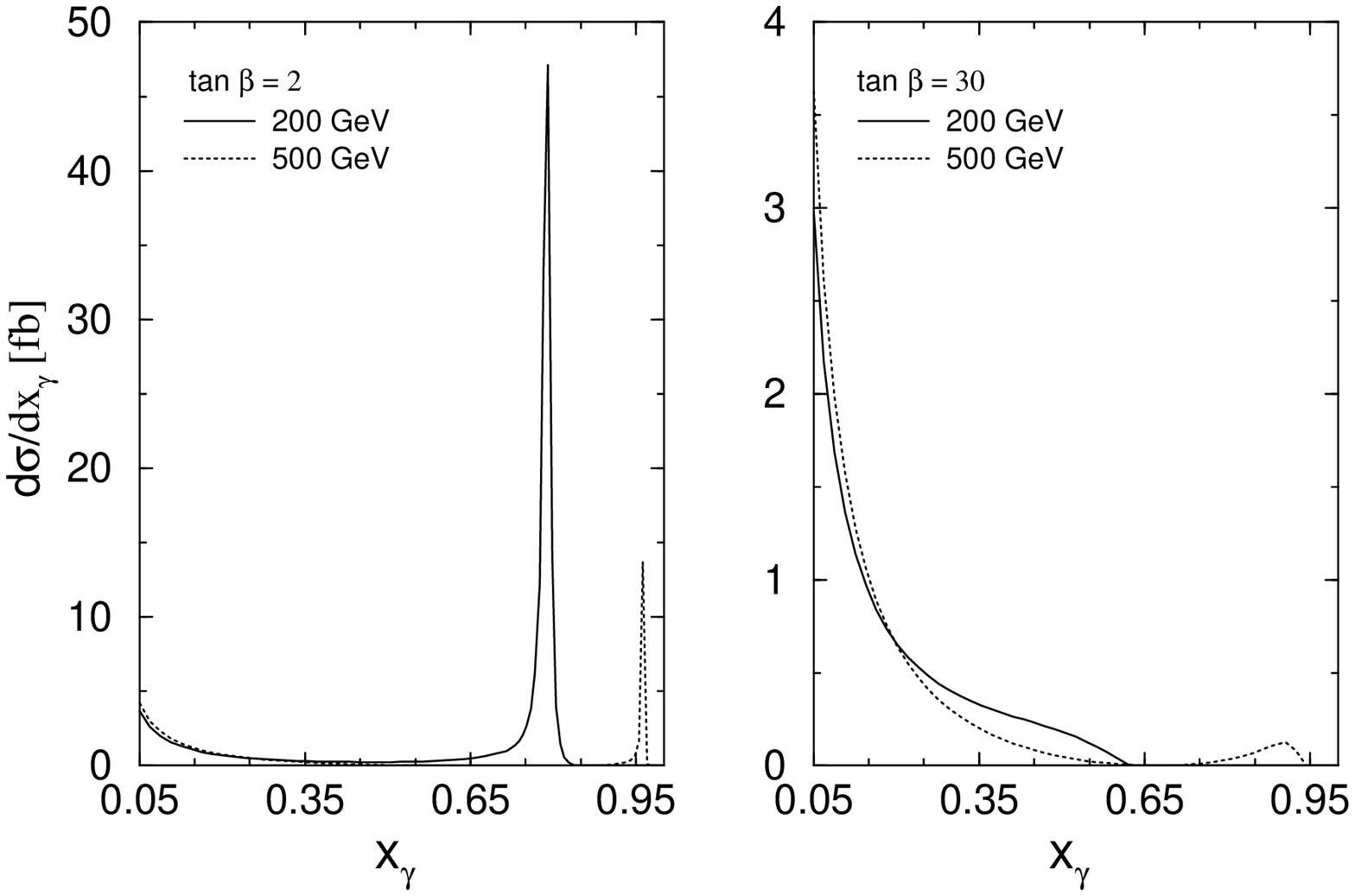,width=16cm,height=10cm}\hss}
\label{fig:fig5}
\end{figure}
\end{center}
\vfill
\begin{center}
{\bf\large Figure 5}
\end{center}

\newpage
\mbox{ }

\begin{center}
\begin{figure}[htb]
\vspace{1.5cm}
\hbox to\textwidth{\hss\epsfig{file=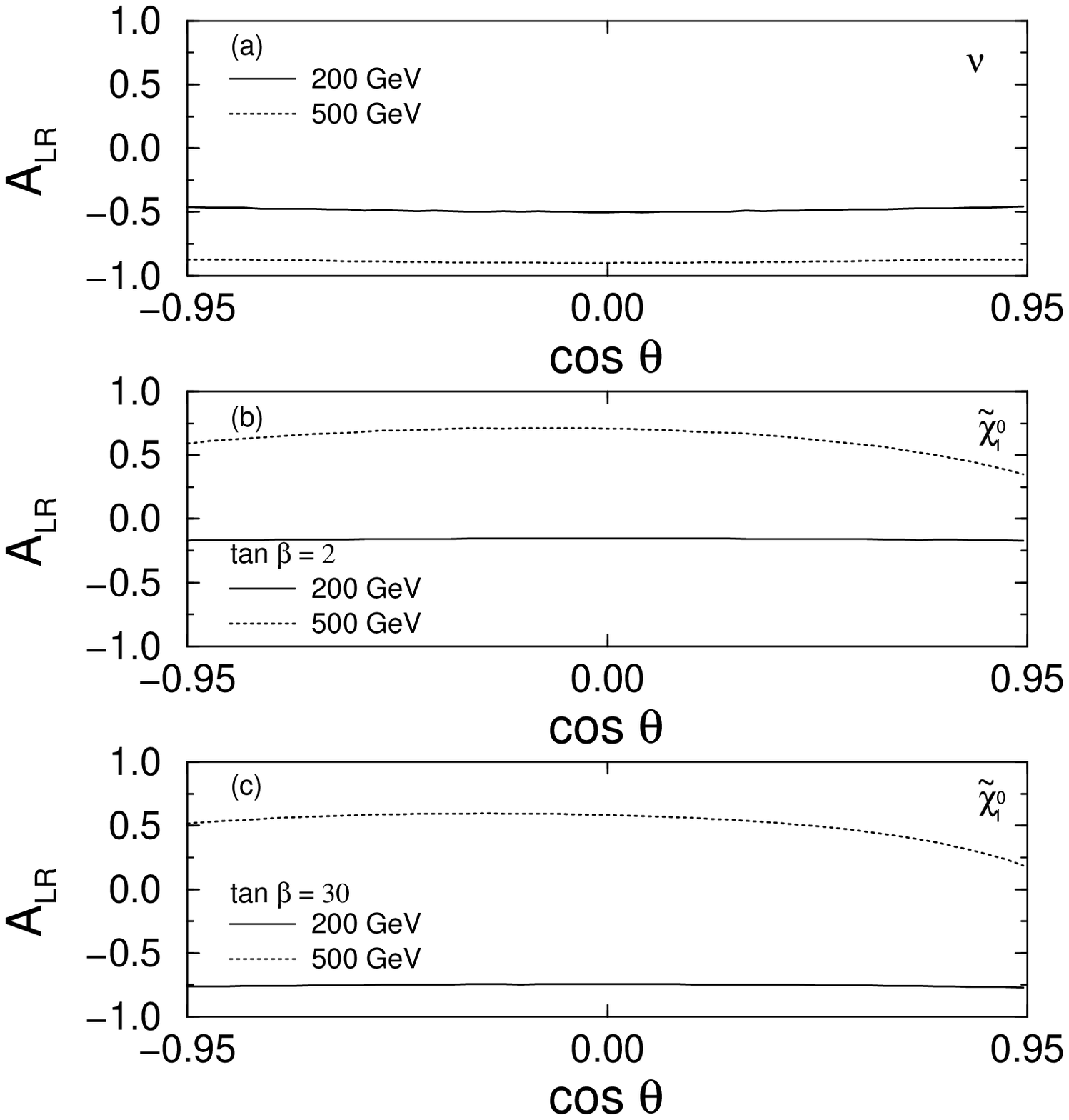,width=7cm,height=15cm}\hss}
\label{fig:fig6}
\end{figure}
\end{center}
\vfill
\begin{center}
{\bf\large Figure 6}
\end{center}

\newpage
\mbox{ }

\begin{center}
\begin{figure}[htb]
\vspace{3.5cm}
\hbox to\textwidth{\hss\epsfig{file=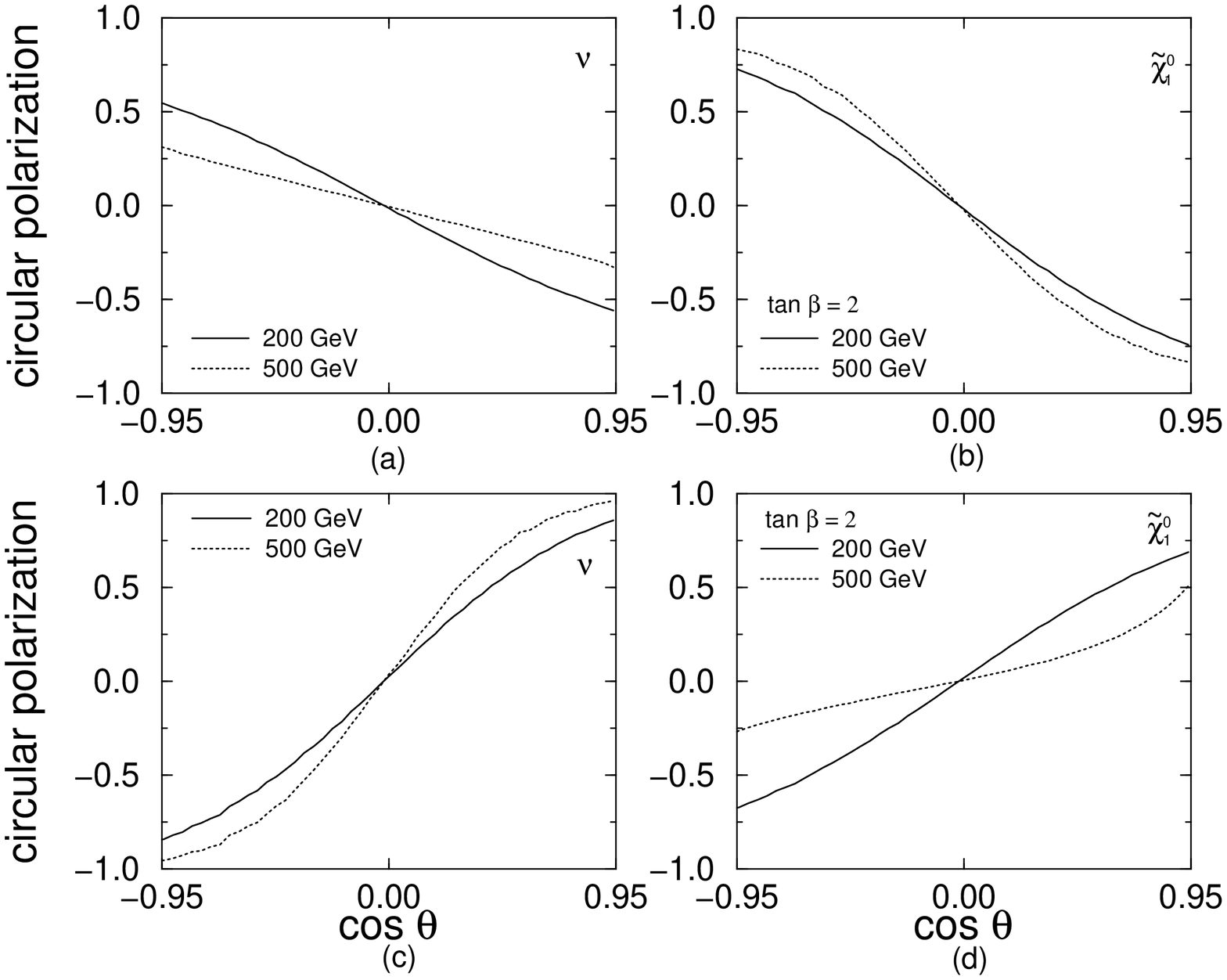,width=13cm,height=10cm}\hss}
\label{fig:fig7}
\end{figure}
\end{center}
\vfill
\begin{center}
{\bf\large Figure 7}
\end{center}

\end{document}